\documentclass[twocolumn,showpacs,preprintnumbers,amsmath,amssymb]{revtex4}
\usepackage{graphicx}% Include figure files
\usepackage{dcolumn}% Align table columns on decimal point
\usepackage{bm}% bold math
\usepackage{latexsym}
\usepackage{amsfonts}
\usepackage{amssymb}
\usepackage{amsmath}
\begin{document}

%\preprint{}

\title{Attractor Solutions in Lorentz Violating Scalar-Vector-Tensor Theory}
\author{Arianto$^{(1,2,3)}$}
\email{feranie@upi.edu}
\author{Freddy P. Zen$^{(1,2)}$}
\email{fpzen@fi.itb.ac.id}
\author{Triyanta$^{(1,2)}$}
\email{triyanta@fi.itb.ac.id}
\author{Bobby E. Gunara$^{(1,2)}$}
\email{bobby@fi.itb.ac.id}
\affiliation{$^{(1)}$Theoretical Physics Lab., THEPI Devision, \\
and \\
$^{(2)}$Indonesia Center for Theoretical and Mathematical Physics (ICTMP)\\
Faculty of Mathematics and Natural Sciences,\\
 Institut Teknologi Bandung,\\
Jl. Ganesha 10 Bandung 40132, INDONESIA. \\
$^{(3)}$Department of Physics, Udayana University\\
Jl. Kampus Bukit Jimbaran Kuta-Bali 80361, INDONESIA.\\
}

%===============================================================%
%************************* ABSTRACT ****************************%
%===============================================================%
\begin{abstract}
We investigate properties of attractors for scalar field in the
Lorentz violating scalar-vector-tensor theory of gravity. In this
framework, both the effective coupling and potential functions
determine the stabilities of the fixed points. In the model, we
consider the constants of slope of the effective coupling and
potential functions which lead to the quadratic effective coupling
vector with the (inverse) power-law potential.  For the case of
purely scalar field, there are only two stable attractor solutions
in the inflationary scenario. In the presence of a barotropic
fluid, the fluid dominated solution is absent. We find two scaling
solutions: the kinetic scaling solution and the scalar
field scaling solutions. We show the stable
attractors in regions of ($\gamma$, $\xi$) parameter space and in
phase plane plot for different qualitative evolutions. From the
standard nucleosynthesis, we derive the constraints for the value
of the coupling parameter.
\end{abstract}

\maketitle

%===============================================================%
%************************ SECTION I ****************************%
%===============================================================%
\section{\label{secI}Introduction}
One of the interesting and widely exploited cosmological models is
the scalar-tensor theory of gravity, the theory of a scalar field
coupled to gravity. A remarkable phenomenon such as inflation, is
usually discussed in the frame of this model. Recently, however, a
scenario of implementing local Lorentz violation in a
gravitational setting is considered to imagine the existence of a
tensor field with a non-vanishing expectation value, and then to
couple this tensor to gravity or matter fields. The vector field
picks out a preferred frame at each point in spacetime, and any
matter fields coupled to it will experience a violation of local
Lorentz invariance~\cite{Colladay:1998, Carroll:1990}. In a
cosmological background, such a vector field acts to rescale the
effective value of Newton's constant~\cite{Carroll:2004ai}.
Moreover, from the study of the spontaneous breaking of Lorentz
symmetry due to a vector field~\cite{Gripaios:2004ms,
Jacobson:2000xp, Kostelecky:1988zi, Mattingly:2005re},  many
current experiments and
observations~\cite{Eling:2003rd,Lim:2004js,Graesser:2005bg,Foster:2005dk,Foster:2006az}
can be explained. In Ref.~\cite{BLi:2008} it has been studied the
late time evolution of the vector field perturbation and its
effects on cosmological observables. They found that the CMB and
matter power spectra are modified.

More recently, a great interest has been devoted to study
cosmological inflation in the framework of Lorentz violating
scalar-vector-tensor theory of gravity~\cite{KS}. They have shown
that the Lorentz violating vector affects the dynamics of the
inflationary model. One of the interesting feature of this
scenario, is the exact Lorentz violating inflationary solutions in
the absence of the inflaton potential. In this case, the inflation
is completely associated with the Lorentz violation. Depending on
the value of the coupling parameter, the three kinds of exact
solutions are found: the power law inflation, de Sitter inflation,
and the superinflation. Moreover, the dynamics of superinflation
in the context of Loop Quantum Cosmology, in which the Friedmann
equation is modified by the presence of inverse volume corrections
has been studied, recently \cite{Copeland:2007}.

The dynamical system of the scalar field with canonical Lagrangian
has been widely studied~\cite{Copeland:1998, Nunes:2001}, among
which the global structure of the phase plane has been
investigated and various critical points and their physical
significances have been identified and manifested. General
properties of attractors for scalar-field dark energy scenarios
which possess cosmological scaling solutions have been studied in
Ref.~\cite{Tsujikawa:2006} (see also \cite{Amendola:2006}).

Cosmological attractor solutions have been found and studied by
several authors for various classes of potentials. The main
purpose of this paper is to show that it is possible to find
attractor solutions in the Lorentz violating scalar-vector-tensor
models in which both the effective coupling function and the
potential function are specified, and their relation matters. In
other words, we will find a class of models in which the dynamics
of the system dependents on the effective coupling function and
the potential function, as well as on their relation. More
specific, we consider the model in which the slope of the
effective coupling vector and the potential in the Lorentz
violation are constants. Then we find the Lorentz violating model
with the quadratic effective coupling vector and the inverse or
power-law-potential. The quintessence scenario with this potential
has been well studied in the conventional universe
\cite{Ratra:1988, Zlatev:1999}. In the case of a tachyon field the
corresponding potential is given by $V=V_0 \phi^{-2}$
\cite{Abramo:2003, Aguirregabiria:2004, Copeland:2005}. The
dynamical attractor of the cosmological system has been employed
to make the late time behaviors of the model insensitive to the
initial condition of the field and thus alleviates the fine tuning
problem. In quintessence models, the dynamical system has tracking
attractor that makes the quintessence evolves by tracking the
equation of state of the background cosmological fluid so as to
alleviating the fine tuning problem ~\cite{Zlatev:1999,
Copeland:1998, Nunes:2001}.

This paper is organized as follows. In Section~\ref{secII}, we set
down the general formalism for the scalar-vector-tensor theory
where the Lorentz symmetry is spontaneously broken due to the
unit-norm vector field. We derive the governing equations of
motion for the canonical Lagrangian of the scalar field. In
Section~\ref{secIII}, we study the attractor solutions for the
purely scalar field.  In Section~\ref{secIV}, we extend our
analysis in the presence of the barotropic fluid. The critical
points of the system and their stability are presented. The final
Section is devoted to the conclusions. In the Appendix, we present
the stability of the fixed point in which both the slope of
effective coupling function and potential function are constant
parameters.

%===============================================================%
%************************ SECTION II ****************************%
%===============================================================%
\section{\label{secII}Lorentz violating scalar-vector-tensor}
In the present section, we develop the general reconstruction
scheme for the scalar-vector-tensor gravitational theory. We will
consider the properties of general four-dimensional universe, i.e.
the universe where the four-dimensional space-time is allowed to
contain any non-gravitational degree of freedom in the framework
of Lorentz violating scalar-tensor-vector theory of gravity. Let
us assume that the Lorentz symmetry is spontaneously broken by
getting the expectation values of a vector field $u^\mu$ as $<0|
u^\mu u_\mu |0> = -1$. The action can be written as the sum of
three distinct parts:
\begin{eqnarray}
     S&=& S_g + S_u + S_{\phi} \ ,
     \label{eq:action}
\end{eqnarray}
where the actions for the tensor field $S_g$, the vector field
$S_u$, and the scalar field $S_{\phi}$, respectively, are given by
\begin{eqnarray}
        S_g &=& \int d^4 x \sqrt{-g}~ {1\over 16\pi G}R  \ ,
    \label{eq:act-grav} \\
  S_u &=& \int d^4 x \sqrt{-g} \left[
  - \beta_1 \nabla^\mu u^\nu \nabla_\mu u_\nu
   -\beta_2 \nabla^\mu u^\nu \nabla_\nu u_\mu  \right. \nonumber\\
 && \left.  -\beta_3 \left( \nabla_\mu u^\mu \right)^2
  -\beta_4 u^\mu u^\nu \nabla_\mu u^\alpha \nabla_\nu u_\alpha
   \right. \nonumber\\
 && \left. + \lambda \left( u^\mu u_\mu +1 \right) \right]  \ ,
  \label{eq:act-VT} \\
  S_\phi &=&  \int d^4 x \sqrt{-g}~ {\cal{L}}_\phi \ .
   \label{eq:act-matter}
\end{eqnarray}
In the above $\beta_i(\phi)$ ($i=1,2,3,4$) are arbitrary
parameters and ${\cal{L}}_{\phi}$ is the Lagrangian density for
scalar field, expressed as a function of the metric $g_{\mu\nu}$
and the scalar field $\phi$. $\lambda$ is a Lagrange multiplier.
Then, the action~(\ref{eq:action}) describes the
scalar-vector-tensor theory. For the time-like vector field, we
impose a constraint
\begin{eqnarray}
  u^\mu u_\mu = -1 .
\end{eqnarray}
Here, we take $u^\mu$ as the dimensionless vector, and
accordingly, $\beta_i$ has the dimension of mass squared. Thus,
${\beta_i}^{1/2}$ gives the mass scale of symmetry breakdown. The
preferred frame determined by the vector $u^\mu$ differs from the
CMB rest frame and the alignment of these frames had been achieved
during the cosmic expansion as is explained in the Appendix of
Ref.~\cite{KS}. In this setup, the preferred frame is selected
through the constrained vector field $u^\mu$ and this leads to
violating the Lorentz symmetry.

For the background solutions, we use the homogeneity and isotropy
of the universe  spacetime
\begin{eqnarray}
ds^2 = - {\mathcal{N}}^2 (t) dt^2 + e^{2\alpha(t)} \delta_{ij}
dx^i dx^j \ ,
\end{eqnarray}
where ${\mathcal{N}}$ is a lapse function. The scale of the
universe is determined by $\alpha$. We take the constraint
\begin{eqnarray}
   u^\mu = \left( {1\over {\mathcal{N}}} , 0 ,0 ,0 \right) \ ,
\end{eqnarray}
where ${\mathcal{N}} =1$ is taken into account after the
variation. Varying the action (\ref{eq:action}) with respect to
$g^{\mu\nu}$, we have field equations
\begin{eqnarray}
   R_{\mu\nu}-{1\over 2}g_{\mu\nu}R = 8\pi G T_{\mu\nu} \ ,
   \label{eq:einstein-eq}
\end{eqnarray}
where $T_{\mu\nu} =T_{\mu\nu}^{(u)} + T_{\mu\nu}^{(\phi)}$ is the
total energy-momentum tensor, $T_{\mu\nu}^{(u)} $ and
$T_{\mu\nu}^{(\phi})$ are the energy-momentum tensors of vector
and scalar fields, respectively, defined by the usual formulae
\begin{eqnarray}
     T_{\mu\nu}^{(k)} = -2\frac{\partial {\cal{L}}^{(k)} }{\partial g^{\mu\nu}
     }+ g_{\mu\nu} {\cal{L}}^{(k)}  , \qquad k=u, {\phi}  \ .
\end{eqnarray}
The time and space components of the total energy-momentum tensor
are given by
\begin{eqnarray}
     T^{0}_{0} = - \rho_u -\rho_{\phi} \ , \qquad    T^{i}_i =  p_u+ p_{\phi} \ ,
     \label{eq:00-ii-compot}
\end{eqnarray}
where the energy density and pressure of the vector field are
given by
\begin{eqnarray}
    && \rho_u =  -3\beta H^2  \ ,
    \label{eq:rho-vf} \\
    &&p_u =  \left(3 + 2{H^{\prime}\over H} + 2{\beta^{\prime}\over \beta} \right)\beta H^2 \ ,
    \label{eq:pres-vf} \\
    && \beta \equiv \beta_1 +3 \beta_2 + \beta_3 \ .
    \label{eq:def-beta}
\end{eqnarray}
Note that $\beta_4$ does not contribute to the background
dynamics. A prime denotes the derivative of any quantities $X$
with respect to $\alpha$. $X^{\prime}$ is then related to its
derivative with respect to $t$ by
$X^{\prime}=(dX/dt)H^{-1}=\dot{X} H^{-1}$ where
$H=d\alpha/dt=\dot{\alpha}$ is the Hubble parameter. From Eqs.
(\ref{eq:rho-vf}) and (\ref{eq:pres-vf}), one obtains the energy
equation for the vector field $u$
\begin{eqnarray}
   {\rho}^{\prime}_u + 3({\rho}_u + p_u)=+3H^2 \beta^{\prime}  \ ,
   \label{eq:eos-vec}
\end{eqnarray}
and for the scalar field
\begin{eqnarray}
    {\rho}^{\prime}_{\phi} + 3({\rho}_{\phi} + p_{\phi})=-3H^2 \beta^{\prime}  \ .
   \label{eq:eos-mat}
\end{eqnarray}
The total energy equation in the presence of both the vector and
the scalar fields is, accordingly,
\begin{equation}
   {\rho}^{\prime} + 3({\rho} + p)=0 \ , \qquad \rho = \rho_u + \rho_{\phi} \ .
   \label{eq:eos-total}
\end{equation}

Substituting Eq.~(\ref{eq:00-ii-compot}) into the Einstein
equations (\ref{eq:einstein-eq}), we obtain two independent
equations, called the Friedmann equations, as follows:
\begin{eqnarray}
    \left( 1 + \frac{1}{8\pi G \beta} \right) H^2&=& {1\over 3\beta} \rho_{\phi}   \ ,
    \label{eq:Friedmann} \\
    \left( 1 + \frac{1}{8\pi G \beta} \right) \left( HH'+H^2\right)&=&-{1\over 6} \left( {\rho_{\phi}\over \beta} + {3p_{\phi}\over \beta} \right) \nonumber\\
    &&- H^2 {\beta'\over \beta} \ .
    \label{eq:Friedmannsec}
\end{eqnarray}
The second term on RHS of Eq.~(\ref{eq:Friedmannsec}) is a
consequence of the coupling vector field as a function of scalar
field. If $\beta_i =0$, thus without the vector field, the above
equations reduce to the conventional ones. And in the case
$\beta_i=const$., the above equations lead to the Friedmann
equations given in Ref. \cite{Carroll:2004ai}.

Let us define the effective coupling function as follows
\begin{eqnarray}
   \bar{\beta} &\equiv&   \beta + {1\over 8\pi G} \ ,
   \label{eq:twopart}
\end{eqnarray}
then Eqs.~(\ref{eq:Friedmann}) and (\ref{eq:Friedmannsec}) can be
rewritten as
\begin{eqnarray}
&&  H^2
  = \frac{1}{3\bar{\beta}} \rho_\phi \ ,
  \label{eq:friedmannnew1}\\
&&  {H'\over H}=- {\bar{\beta}'\over \bar{\beta}} - {3\over 2 }
\gamma_{\phi} \ .
     \label{eq:friedmannnew2}
\end{eqnarray}
where the equation of state for the scalar field is defined by
$\gamma_{\phi} = 1+p_{\phi}/ \rho_{\phi}$.

For a given scalar field Lagrangian with the FRW background, we
can obtain the equations of motion for a scalar field by using
Eq.~(\ref{eq:eos-mat}). Let us consider the Lagrangian density of
a scalar field $\phi$ with a potential $V(\phi)$ in
Eq.~(\ref{eq:action}):
\begin{eqnarray}
{\cal{L}}_{\phi}= -{1\over 2}(\nabla \phi)^2 - V(\phi) \ ,
\end{eqnarray}
where $(\nabla \phi)^2=g^{\mu\nu}
\partial_{\mu}\phi\partial_{\nu}\phi$. For the homogeneous field,
the energy density $\rho_{\phi}$ and the pressure $p_{\phi}$ of
the scalar field may be found as follows
\begin{eqnarray}
&& {\rho_\phi \over V} =  \left( {6\bar{\beta}\over 6\bar{\beta} -
\phi^{\prime 2} }\right) \ ,
\label{eq:infrho} \\
&&{p_{\phi}\over V}=  -2\left( {3\bar{\beta}-\phi^{\prime 2} \over
6\bar{\beta} -  \phi^{\prime 2}}\right)  \ ,
\label{eq:infp}\\
&&\gamma_{\phi}={\phi^{\prime 2}\over 3\bar{\beta}} \ .
\label{scl-eos}
\end{eqnarray}
Substituting Eq.~(\ref{eq:infrho}) into
Eq.~(\ref{eq:friedmannnew1}) and also substituting
Eq.~(\ref{scl-eos}) into Eq.~(\ref{eq:friedmannnew2}), the
Friedmann equations lead to
\begin{eqnarray}
&& H^2\left( 1 -  {\phi^{\prime 2}\over6\bar{\beta}} \right)
={1\over 3\bar{\beta}}V \ ,
  \label{eq:friedmannnew2-1}\\
&&{H'\over H}=- {\bar{\beta}'\over \bar{\beta}} - {1\over 2 }
{\phi'^2\over \bar{\beta} }  \ .
     \label{eq:friedmannnew2-2}
\end{eqnarray}
Now differentiating Eq.~(\ref{eq:infrho}) with respect to $\alpha$
and using Eq.~(\ref{eq:eos-mat}), we obtain a dynamical equation
for the scalar field $\phi$,
\begin{eqnarray}
   \phi'' =- \left( \frac{H'}{H} + 3\right)\phi' -  \frac{V_{,\phi}}{H^2}
             - 3\bar{\beta}_{,\phi}\ ,
 \label{eq:03-1}
\end{eqnarray}
which is subject to the Friedmann constraint given by
Eq.~(\ref{eq:friedmannnew2-1}).

%===============================================================%
%************************ SECTION III **************************%
%===============================================================%
\section{\label{secIII}Attractor Solutions for Purely Scalar Field}
Equations~(\ref{eq:friedmannnew2-1})--(\ref{eq:03-1}) are the
governing equations of motion which we will use to study dynamical
attractor for purely scalar field. We introduce the following
dimensionless variables~\cite{Arianto:2007}:
\begin{eqnarray}
    &&x^2\equiv{\phi'^2 \over 6\bar{\beta}} \ , \qquad \quad y^2 \equiv {V\over 3H^2\bar{\beta}} \ ,
    \label{def-xy}\\
    && \lambda_1 \equiv -{\bar{\beta}_{,\phi}\over \sqrt{\bar{\beta}}} \ , \qquad \lambda_2 \equiv - \sqrt{\bar{\beta}}{V_{,\phi}\over V} \ ,
    \label{def-lambda}\\
    && \Gamma_1 \equiv \frac{\bar{\beta} \bar{\beta}_{,\phi\phi}}{\bar{\beta}_{,\phi}^2} \ , \qquad \Gamma_2 \equiv \frac{V V_{,\phi\phi}}{V_{,\phi}^2} +{1\over 2}{\bar{\beta}_{,\phi}/\bar{\beta}\over V_{,\phi}/V}\ ,
    \label{def-gamma}
\end{eqnarray}
and, accordingly, the governing equations of motion could be
reexpressed as the following system of equations:
\begin{eqnarray}
    &&x'= -3x(1-x^2)+\sqrt{{3\over 2}}\left(\lambda_1 + \lambda_2  \right)y^2 \ ,
    \label{auto-x}\\
    && y' = \left[ 3x -\sqrt{{3\over 2}}\left(\lambda_1 + \lambda_2 \right)\right]x y \ ,
    \label{auto-y}\\
    && \lambda_1^{\prime} =-\sqrt{6}\lambda_1^2  \left(\Gamma_1 -{1\over 2}\right)x  \ ,
    \label{auto-L1}\\
    &&\lambda_2^{\prime} =-\sqrt{6}\lambda_2^2   \left( \Gamma_2 -1 \right) x \ ,
    \label{auto-L2}
\end{eqnarray}
where a prime denotes a derivative with respect to the logarithm
of the scale factor, $\alpha=\ln a$.

In general, the parameters $\lambda_1$, $\lambda_2$, $\Gamma_1$
and $\Gamma_2$ are  variables dependent on $\phi$ and completely
associated with the Lorentz violation. In particular, $\lambda_1$
and $\Gamma_1$ are purely Lorentz violation parameters and
$\Gamma_1$ can be written as a function of $\lambda_1$ in this
case. By definition (\ref{def-gamma}), $\Gamma_2$ can be written
as a function of $\lambda_1$ and $\lambda_2$. Thus, in order to
construct viable Lorentz violation model, we require that the
coupling function $\bar{\beta}$ and the potential function $V$
should satisfy the condition $\Gamma_1 >1/2$ and $\Gamma_2 >1$,
respectively. In this paper, we want to discuss the phase space,
then we need certain constraints on the coupling function and
potential function. In particular, we study the case of constants
$\Gamma_1$ and $\Gamma_2$. We take the form of $\Gamma_1$ and
$\Gamma_2$ as $\Gamma_1 =1/2$ and $\Gamma_2 =1$, respectively.
Equations (\ref{auto-L1}) and (\ref{auto-L2}) imply that
$\lambda_1$ and $\lambda_2$ are nearly constants in this case.

Integrating Eq.~(\ref{def-gamma}) with respect to $\phi$, we
obtain a Lorentz violating model
\begin{eqnarray}
    \bar{\beta}(\phi) = \xi \phi^2 \ , \qquad V(\phi)=V_0 \phi^{n} \ ,
    \label{model}
\end{eqnarray}
where $\xi$, $n$ and $V_0$ are parameters. Hence,
\begin{eqnarray}
    \lambda_1 = -2\sqrt{\xi} \ , \qquad \lambda_2 = {n\over 2} \lambda_1 \ .
\end{eqnarray}
In the following, we study the case power-law potential $n>0$.

From the above equation, we require $\xi>0$. Then, the dynamical
system of equations (\ref{auto-x})--(\ref{auto-L2}) can be
rewritten as an autonomous system:
\begin{eqnarray}
    &&x'= -3x(1-x^2)-( n+ 2)\sqrt{{3\xi\over 2}} y^2 \ ,
    \label{autosys-x}\\
    && y' =  \left[ 3x + ( n+ 2)\sqrt{{3\xi\over 2}}\right]x y \ .
    \label{autosys-y}
\end{eqnarray}
The equation (\ref{eq:friedmannnew2-1}) leads to the constraint
\begin{eqnarray}
    x^2 + y^2 =1 \ .
    \label{Fried-const-1}
\end{eqnarray}
In term of the new variable, the equation
(\ref{eq:friedmannnew2-2}) reads
\begin{eqnarray}
  {H'\over H}= -3x^2 - 2\sqrt{6\xi}  x \ .
     \label{selfsim}
\end{eqnarray}
Integrating this equation with respect to $\alpha$ one shows that
all critical points, where $x$ is a non-zero constant, correspond
to an evolution of the Hubble parameter $H$ given by $H \propto
e^{-\alpha/p}$, where
\begin{eqnarray}
  p \equiv {1\over 3x^2 +2 \sqrt{6\xi}  x} \ .
  \label{defpowerlaw}
\end{eqnarray}
\begin{figure}[h]% fig.1
%\begin{center}
\includegraphics[height=7cm, width=8cm]{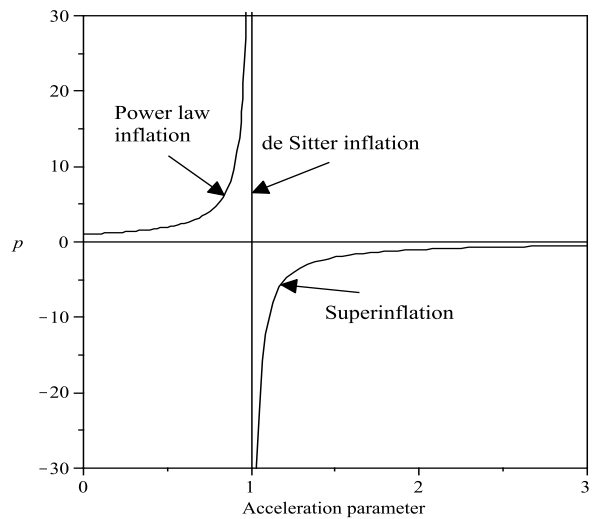}
%\end{center}
\caption{There are three possibilities of the inflationary
solutions in the Lorentz violating scalar-vector-tensor theory:
the power-law inflation, the de Sitter inflation and the
superinflation. } \label{fig:typeinflation}
\end{figure}

An inflationary phase is one wherein the universe undergoes an
accelerating expansion, i.e., the scale factor satisfies $\ddot{a}
> 0$. Inflation ends when this condition is violated. In the frame
of Lorentz violating scalar-vector-tensor, the condition for the
accelerating universe is
\begin{eqnarray}
 \sqrt{\xi}  x <  {1\over 6}\sqrt{{3\over 2}} \left( 1 -{3\over 2}\gamma_{\phi} \right) \ ,
  \label{acc:cond1}
\end{eqnarray}
where $\gamma_\phi$ is an equation of state for the scalar field,
$\gamma_\phi=1+\omega_\phi$. When the condition (\ref{acc:cond1})
is satisfied, we find three types of the inflationary solution
given by Eq.~(\ref{defpowerlaw}): the  power-law inflation
($p>0$), the  de Sitter inflation ($1/p=0$) and the superinflation
($p=-|p|<0$), depending on the values of $\xi$. We will show that
there exists a stable attractor for this three types of the
inflationary solution in the framework of Lorentz violating
scalar-vector-tensor theory of gravity. In the conventional case,
$\xi \rightarrow 0$, we are only left with a power-law inflation
$p=1/3x^2$, where the condition for the accelerated universe is,
accordingly, $\gamma_{\phi}<2/3$. In Fig.~\ref{fig:typeinflation}, we show
three possibilities of the inflationary solutions. Here we define
an acceleration parameter,
\begin{equation}
    \bar{q}\equiv 1+ \frac{\ddot{\alpha}}{\dot{\alpha}^2}=1-{3\over 2}\gamma_{\phi}-2 \sqrt{6\xi} \ ,
\end{equation}
that simplifies
\begin{eqnarray}
  p ={1\over 1-\bar{q}} \ .
  \label{plwithmatter-simplify-1}
\end{eqnarray}
Then, the condition for the accelerating universe
(\ref{acc:cond1}) becomes $\bar{q}>0$. It can be shown that the
inflationary (accelerated) solutions are classified into three
cases (see Fig.~\ref{fig:typeinflation}):  a)~$0 < \bar{q} <1$, a
solution in which $p>0$. In this case we have a power law
inflation where $a(t) \sim t^p$.   b)~$\bar{q} =1$, $1/p =0$. This
is a de Sitter solution $a(t) \sim e^{Ht}$. c)~$ \bar{q}>1$. In
this case, $p\equiv -|p| <0$. Hence, the solution becomes $a(t)
\sim (-t)^{-|p|}$, $t < 0 $. Thus, this solution represents the
super-inflationary universe.

In order to study the stability of the critical points, using the
constraint Eq.~(\ref{Fried-const-1}), we first reduce
Eqs.~(\ref{autosys-x}) and (\ref{autosys-y}) to a one dimensional
equation,
\begin{equation}
    x'= -3\left[x+(n+ 2) \sqrt{ {\xi\over 6}} \right](1-x^2) \ .
    \label{auto-x-1}
\end{equation}
If we linearize the system about the critical points $x
\rightarrow x_c + u$, we could readily write the first-order
perturbation equation as
\begin{equation}
    u'= \left[2x\left(3x +( n+2)\sqrt{{3\xi\over 2}} \right) -3(1-x^2)\right]u \ ,
    \label{auto-x-1-perturb}
\end{equation}
which yields one eigenvalue.

Depending on the values of $\xi$ and $n$, we find three critical
points where $x'$ and $y'$ vanish:
\begin{itemize}
    \item Point $(A_{\pm})$:\\
    ($x_c,y_c$)=($\pm1,0$) is a kinetic dominated solution.
    \item Point $(B)$:\\
    ($x_c,y_c$)=($-(n+2)\sqrt{\xi/6}$, $\sqrt{1-(n+2)^2\xi/6}$) is a potential-kinetic dominated solution.
\end{itemize}

Notice that the above critical points depend on the values of the
effective coupling parameter $\xi$ and the power of the potential
$n$. For the positive values of $\xi$ and $n$, the linear
perturbation shows that the point ($A_{+}$) is always unstable,
whereas the point ($A_{-}$) is stable for $\xi>6/(n+2)^2$. Thus,
we only left with one possibility attractor solution in the
kinetic dominated solution. The potential-kinetic dominated
solution, point ($B$), is stable for $0<\xi<6/(n+2)^2$. In this
case, the point ($A_{-}$) is unstable.

For the point ($A_{-}$), the universe accelerates for  $\xi>1/6$.
Combining the stability of the critical point and the condition
for the accelerating universe, the inflationary attractor
solutions are given by $0<n\leq 4$, $\xi>6/(n+2)^2$ and $n>4$,
$\xi>1/6$. Then, we find that the three types of inflationary
(accelerated) solutions depend on the potential parameter and the
effective coupling parameter.
\begin{itemize}
    \item The power-law inflationary attractor:
\begin{eqnarray}
    && 2<n\leq 4 \ , \qquad \frac{6}{(n+2)^2} <\xi<\frac{3}{8} \ ,\\
    && n>4 \ , \qquad\qquad 1/6<\xi<3/8 \ .
\end{eqnarray}
    \item The de Sitter inflationary attractor:
\begin{eqnarray}
    n > 2 \ , \qquad \xi=\frac{3}{8} \ .
\end{eqnarray}
    \item The super-inflationary attractor:
\begin{eqnarray}
    && 0<n\leq 2 \ , \qquad \xi>\frac{6}{(n+2)^2} \ , \\
    && n>2 \ , \qquad\qquad \xi> \frac{3}{8} \ .
\end{eqnarray}
\end{itemize}

In the case potential-kinetic dominated solution, the accelerated
expansions of the universe are given by $0<n\leq2$, $\xi>0$
and $n > 2$, $0<\xi<2/(n^2-4)$. Then we find
\begin{itemize}
    \item The power-law inflationary attractor:
\begin{eqnarray}
   &2<n\leq 4& \ , \qquad 0 <\xi<\frac{6}{(n+2)^2} \ , \\
   &n >4& \ , \qquad 0 <\xi<\frac{2}{(n^2-4)} \ .
\end{eqnarray}
   \item The de Sitter inflationary attractor:
\begin{eqnarray}
   \quad n=2 \ , \qquad   0<\xi <3/8 \ .
\end{eqnarray}
    \item The super-inflationary attractor:
\begin{eqnarray}
    0<n< 2 \ , \qquad 0<\xi<\frac{6}{(n+2)^2} \ .
\end{eqnarray}
\end{itemize}

In summary, the effective coupling parameter and the potential
model determine the stabilities of the critical points and the
inflationary solutions. There are two possibilities of attractor
solutions given by the critical point ($A_{-}$) and ($B$).

%===============================================================%
%************************ SECTION IV ***************************%
%===============================================================%
\section{\label{secIV} Attractor solutions in the presence of barotropic fluid}
For a realistic model, we consider the effect of an additional
component. We carry out cosmological dynamics of the scalar field
$\phi$ in the presence of a barotropic fluid whose equation of
state is given by $p_{\gamma} = (\gamma - 1)\rho_{\gamma}$, where
$\gamma$ is an adiabatic index, $0\leq\gamma\leq 2$. We assume
that there is a barotropic fluid, not explicitly coupled to the
scalar field and the vector field. Then, the total energy-momentum
tensor in Eq.~(\ref{eq:einstein-eq}) for this case is $T_{\mu\nu}
=T_{\mu\nu}^{(u)} + T_{\mu\nu}^{(\phi)} +T_{\mu\nu}^{(\gamma)}$,
where
\begin{eqnarray}
    T_{\mu\nu}^{(\gamma)} &=& (\rho_\gamma + p_\gamma)n_\mu n_\nu + p_\gamma g_{\mu\nu}
    \ ,
    \label{eq:emmatter}
\end{eqnarray}
is the energy-momentum tensor of the matter field. Here $n^\mu$ is
the four velocity. The time and space components of the Einstein
equations (\ref{eq:einstein-eq}) yield
\begin{equation}
    3H^2= 8\pi G \left({\rho}_u +{\rho}_\phi +{\rho}_\gamma \right)  \ ,
    \label{eq:Friedmannmatter}
\end{equation}
and
\begin{equation}
    2H^{\prime}H= -8\pi G \left({\rho}_u +{p}_u +{\rho}_\phi+{p}_\phi +{\rho}_\gamma+{p}_\gamma
    \right)\ .
    \label{eq:Raychaudhurimatter}
\end{equation}
Substituting equations (\ref{eq:rho-vf}) and (\ref{eq:infrho}) for
the energy density of vector field and scalar field into equation
(\ref{eq:Friedmannmatter}), respectively, one finds
\begin{equation}
H^2  =  {1\over 3\bar{\beta}}\left({V+\rho_{\gamma}\over 1 -
{\phi^{\prime 2}\over6\bar{\beta}}  }\right)  \ .
\label{friedmannplus}
\end{equation}
The evolution equation for a barotropic fluid is
\begin{eqnarray}
   \rho^{\prime}_{\gamma} =  -3\gamma\rho_{\gamma} \ .
   \label{eos:perfectfluid}
\end{eqnarray}
The scalar field obeys the same equation of motion,
Eq.~(\ref{eq:03-1}). The second Friedmann equation,
Eq.~(\ref{eq:Raychaudhurimatter}), becomes
\begin{eqnarray}
{H'\over H}=- {\bar{\beta}'\over \bar{\beta}} - {1\over 2 }
{\phi'^2\over \bar{\beta} } - \gamma{ \rho_{\gamma}\over
2H^2\bar{\beta}} \ .
\label{friedmannn2plus}
\end{eqnarray}
Equation~(\ref{eq:03-1}) together with
Eqs.~(\ref{friedmannplus})--(\ref{friedmannn2plus}) are the
governing equations of motion which will be used to study
dynamical attractor for a scalar field in the presence of the
barotropic fluid.

For the cases of the constants $\lambda_1$ and $\lambda_2$, the
governing equations can be written as the two-dimensional
autonomous system:
\begin{eqnarray}
    x'&=& {3\over 2}x \left[\gamma (1-x^2-y^2)+2x^2  \right]-3x \nonumber\\
    &&+\sqrt{{3\over 2}} \left[ \lambda_1 \left( 1- x^2  \right)+ \lambda_2 y^2\right] \ ,
    \label{autobarotropic-x}\\
     y' &=& {3\over 2}y \left[\gamma (1-x^2-y^2)+2x^2 \right]\nonumber\\
    && -\sqrt{{3\over 2}} (\lambda_1+\lambda_2) x y \ .
    \label{autobarotropic-y}
\end{eqnarray}%
The system of equations are symmetric under the reflection $(x, y)
\rightarrow (x,-y)$. In what follows, we will restrict our
discussion to the existence and stability of critical points to
the upper half plane $y \geq 0$. In the case $\beta_i=$const.,
$\lambda_1\rightarrow 0$, the scalar field dynamics in the Lorentz
violating scalar-vector-tensor theories is then reduced to the
scalar field dynamics in the conventional one. But, the effective
gravitational constant is rescaled by Eq.~({\ref{eq:twopart}}). In
this case, the cosmological attractor solutions can be studied by
a scalar exponential potential of the form $V(\phi) =
V_0\exp(-\lambda_2\phi/\sqrt{\bar{\beta}})$ where
$\bar{\beta}=$const. This exponential potential gives rise to
scaling solutions for the scalar field \cite{Copeland:1998}.

Also, the Friedmann constraint, Eq.~(\ref{friedmannplus}), becomes
\begin{eqnarray}
    \Omega_{\phi} + {\rho_{\gamma}\over 3H^2 \bar{\beta}} =1 \ ,
    \label{Fried-const-1}
\end{eqnarray}
where the contribution of the scalar field to the total energy
density is
\begin{eqnarray}
    \Omega_{\phi}= {\rho_{\phi}\over 3H^2 \bar{\beta}} =x^2 + y^2 \ .
    \label{Fried-const-1}
\end{eqnarray}
In term of the new variable we find
\begin{eqnarray}
  {H'\over H}= -{3\over 2} \gamma_{eff} \ ,
     \label{selfsim}
\end{eqnarray}
where we have defined the effective equation of state of the
universe:
\begin{eqnarray}
  \gamma_{eff} &\equiv&  \gamma + \Omega_{\phi}\left( \gamma_{\phi}-\gamma \right)-2\sqrt{{2\over 3}}~\lambda_1 x \ .
  \label{def:eoseff}
\end{eqnarray}
When $x$ is a non zero constant, Eq.~(\ref{selfsim}) corresponds
to an evolution of the Hubble parameter given by
\begin{eqnarray}
   H \propto e^{-\alpha /p}  \ ,
\end{eqnarray}
where
\begin{eqnarray}
  p \equiv {1\over {3\over 2} \gamma_{eff}} \ .
  \label{plwithmatter}
\end{eqnarray}

In the presence of barotropic fluid, the condition for the
accelerating universe is
\begin{eqnarray}
  \lambda_1 x > {1\over 2} \sqrt{{3\over 2}} \left( \gamma + \Omega_{\phi}\left( \gamma_{\phi}-\gamma \right) -{2\over 3} \right) \ ,
  \label{acc:cond2}
\end{eqnarray}
The equation (\ref{plwithmatter}) can be simplified by
\begin{eqnarray}
  p ={1\over 1-\bar{q}} \ ,
  \label{plwithmatter-simplify}
\end{eqnarray}
where we have defined an acceleration parameter $\bar{q}$,
\begin{eqnarray}
 \bar{q}=1-{3\over 2} \gamma_{eff}  \ .
  \label{acc:parameter}
\end{eqnarray}
Thus the three types of inflation are also possible in the
presence of the barotropic fluid.

\subsection{Stability of the fixed points}
The critical points $(x_c, y_c)$ are obtained by imposing the
conditions $x^{\prime}=0$ and $y^{\prime}=0$. Substituting linear
perturbation $x \rightarrow x_c+u$ and $y \rightarrow y_c+v$ about
the critical points into Eqs.~(\ref{autobarotropic-x}) and
(\ref{autobarotropic-y}), we obtain, to first-order in the
perturbation, the equations of motion
\begin{eqnarray}
\label{matrix} \left(
  \begin{array}{c}
    u^{\prime} \\
    v^{\prime} \\
  \end{array}
\right)= M\left(
  \begin{array}{c}
    u \\
    v \\
  \end{array}
\right) \ .
\end{eqnarray}
In what follows we clarify the properties of the five critical
points given in Table~\ref{tab:table1} (see Appendix). We analyze
the stability of the  critical points with the background
barotropic fluid $0\leq \gamma \leq 2$. The stable solutions of
these critical points are
\vskip 0.5cm \noindent
a)\ \ Two kinetic dominated solutions \\
For the point ($A_{+}$), the eigenvalues are
\begin{eqnarray}
    m_1 = 3 + (n+2)\sqrt{{3\xi \over 2}} \ , \quad m_2=3(2-\gamma)+2\sqrt{6\xi} \ .
\end{eqnarray}
Thus, the critical point ($A_{+}$) is always unstable for $n>0$.

For the critical point  ($A_{-}$), the eigenvalues are
\begin{eqnarray}
    m_1 = 3 - (n+2)\sqrt{{3\xi \over 2}} \ , \quad m_2=3(2-\gamma)-2\sqrt{6\xi} \ .
\end{eqnarray}
The stable solutions for the kinetic dominated solution ($A_{-}$)
are
\begin{eqnarray}
    0 \leq \gamma  \leq \frac{2n}{n+2} \ ,  \qquad  \xi>\frac{3(2-\gamma)^2}{8}\
    ,
\end{eqnarray}
and
    \begin{eqnarray}
   \frac{2n}{n+2} <\gamma \leq 2\ , \qquad \xi> {6 \over (n+2)^2} \ .
    \end{eqnarray}

In this kinetic dominated solution, we have the following
relations:
\begin{eqnarray}
  && \Omega_{\phi}=1 \ , \quad \gamma_{\phi} =2\ , \quad \gamma_{eff} = 2+4\sqrt{{2\xi\over 3}} \ ,
   \label{OmegaA} \\
 && {1\over p_A}=1-\bar{q}_A \ , \quad \bar{q}_A= -2+2\sqrt{6\xi} \ ,
   \label{pA}
\end{eqnarray}
\vskip 0.5cm \noindent
b)\ \ Scalar field dominated solution, point ($B$)\\
The eigenvalues are
\begin{eqnarray}
    m_1 = -3 + {(n+2)^2\over 2}\xi \ , \quad m_2=-3\gamma+n(n+2)\xi \ .
\end{eqnarray}
The stable solutions for the scalar field dominated solution  are \\
\begin{eqnarray}
    0 <\gamma < \frac{2n}{n+2} \ ,  \qquad  0<\xi<\frac{3\gamma}{n(n+2)}\
    ,
\end{eqnarray}
and
    \begin{eqnarray}
   \frac{2n}{n+2} \leq \gamma \leq 2\ , \qquad 0<\xi< {6 \over (n+2)^2} \ .
    \end{eqnarray}

The critical point exists for  $0<\xi< 6/(n+2)^2$. In this case,
we find the following relations:
\begin{eqnarray}
  && \Omega_{\phi}=1,~\gamma_{\phi} ={1\over 3}(n+2)^2\xi,~  \gamma_{eff} = {1\over 3} (n^2-4)\xi\ ,
   \label{OmegaB} \\
 && {1\over p_B}=1-\bar{q}_B \ , \quad \bar{q}_B= 1-{1\over 2}(n^2-4)\xi \
 .
   \label{pB}
\end{eqnarray}
The universe accelerates for $0<n \leq 2$, $\xi>0$ and $n>2$,
$0<\xi<2/(n^2-4)$.

\vskip 0.5cm \noindent
c)\ \ Kinetic scaling solution, point ($C$) \\
The eigenvalues are
\begin{eqnarray}
    m_1 =  \frac{3\gamma}{2}-\frac{2n\xi}{2-\gamma} \ , \quad m_2= -\frac{3(2-\gamma)}{2}+\frac{4\xi}{(2-\gamma)}\ .
\end{eqnarray}
The stable solutions for the kinetic scaling solution are
\begin{eqnarray}
0 \leq \gamma  < \frac{2n}{n+2} \ ,  \qquad
\frac{3\gamma(2-\gamma)}{4n}<\xi<\frac{3(2-\gamma)^2}{8}\ .
\end{eqnarray}
We obtain the following relations:
\begin{eqnarray}
  && \Omega_{\phi}={8\xi \over 3(2-\gamma)^2} \ ,
   \label{OmegaC} \\
  && \gamma_{\phi} =2 \ , \quad \gamma_{eff} = \gamma-{8\xi \over 3(2-\gamma)}  \ ,
  \label{gamma:C}\\
 && {1\over p_C}=1-\bar{q}_C \ , \quad \bar{q}_C= 1-{3\gamma \over 2}+ {4\xi\over (2-\gamma)} \ .
   \label{pC}
\end{eqnarray}
The universe accelerates for $0\leq \gamma \leq 2/3$, $\xi>0$ and
$2/3<\gamma<2$, $\xi>(3\gamma-2)(2-\gamma)/8$.

Note that the inflationary attractor solution dose not exist for
$\gamma=4/3$. However, in the background matter ($\gamma=1$), the
de Sitter inflationary attractor corresponds to the parameters,
$n>3$, $0<\xi$. the superinflationary attractor does not exist in
this case and the power-law inflationary attractor corresponds to
the parameters, ~$2<n\leq 6$, $3/4n<\xi<3/8$ and $n>6$,
$1/8<\xi<3/8$.

\vskip 0.5cm \noindent
d)\ \ Scalar field scaling solution, point ($D$)\\
The eigenvalues are
\begin{eqnarray}
    m_{1,2} ={T\over 2}\left(1 \pm \sqrt{1-{4D\over T^2}}\right)\
    ,
\end{eqnarray}
where
\begin{eqnarray}
    &&T=-3\left(1 - {(n+2)\gamma \over 2n} \right) \ ,\\
    &&D=\frac{3\left[4n\xi-3\gamma(2-\gamma) \right]\left[3\gamma-(n+2)n\xi
    \right]}{2n^2\xi} \ .
\end{eqnarray}
The stable solutions for the scalar field scaling solution are
given by
\begin{eqnarray}
0 < \gamma  < \frac{2n}{n+2} \ ,  \qquad
\frac{3\gamma}{n(n+2)}\leq \xi \leq \frac{3\gamma(2-\gamma)}{4n}\
.
\end{eqnarray}

The critical point exists for $0<\gamma<2$, $0<\xi<
3\gamma(2-\gamma)/4n$. We find the following relations:
\begin{eqnarray}
  && \Omega_{\phi}={3\gamma\over n^2\xi}\left(1-{2n\xi\over 3\gamma} \right) \ ,
   \label{OmegaD} \\
 &&  \gamma_{\phi} = \frac{3\gamma^2}{3\gamma-2n\xi}  \ , \quad \gamma_{eff} = {\gamma(n-2) \over n} \ ,
   \label{gammaphiD}\\
&& {1\over p_D}=1-\bar{q}_D \ , \quad \bar{q}_D=
1-{3\gamma(n-2)\over 2n} \ ,
   \label{pD}
\end{eqnarray}
and the universe accelerates for $0<n<3$, $0<\gamma<2$ and $n \geq
3$, $0<\gamma<2n/3(n-2)$.

From the above analyzes of the critical points, when one of the
critical point is stable, then the other critical points are
unstable or saddle depending on the values of $\gamma$ and $\xi$.

\subsection{Phase-space diagrams}
\begin{figure}[h]% fig.2
%\begin{center}
\includegraphics[height=7cm, width=8cm]{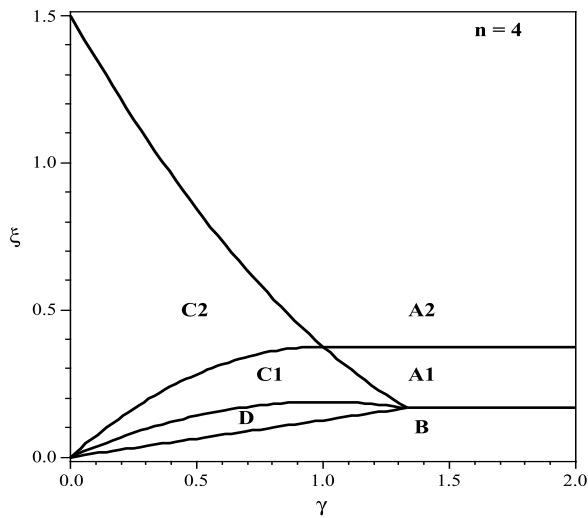}
%\end{center}
\caption{Region of ($\gamma$, $\xi$) parameter space for the
potential model  $V(\phi)=V_0\phi^{4}$.   } \label{fig:phasespace}
\end{figure}

\begin{figure}[h]% fig.3
%\begin{center}
\includegraphics[height=5cm, width=8cm]{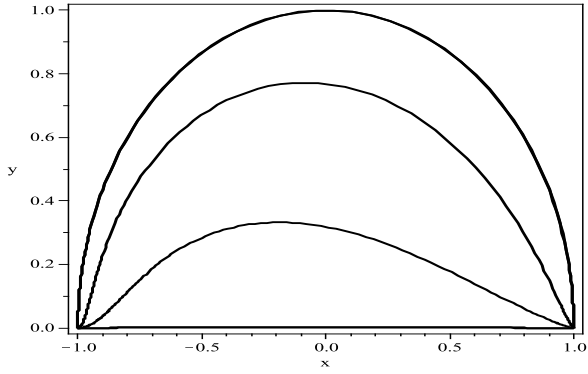}
%\end{center}
\caption{The phase plane for the potential model $V=V_0\phi^{4}$,
$\gamma=1$, and $\xi=1$. The attractor is a kinetic dominated
solution with $x=-1$, $y=0$, where $\Omega_{\phi}=1$,
$\gamma_{\phi}=2$, $\gamma_{eff}=2(1-2\sqrt{2/3})$ and
$\bar{q}=2(\sqrt{6}-1)$. } \label{fig:kineticdom}
\end{figure}

\begin{figure}[h]% fig.4
%\begin{center}
\includegraphics[height=5cm, width=8cm]{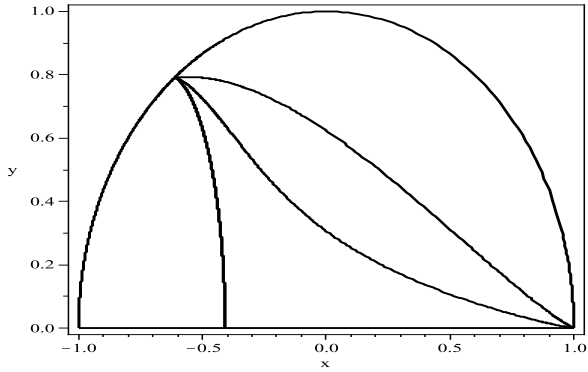}
%\end{center}
\caption{The phase plane for the potential model $V=V_0\phi^4$,
$\gamma=1$, and $\xi=1/16$. The attractor is a scalar field
dominated solution with $x=-\sqrt{3/8}$, $y=\sqrt{5/8}$, where
$\Omega_{\phi}=1$, $\gamma_{\phi}=3/4$, $\gamma_{eff}=1/4$ and
$\bar{q}=5/8$.} \label{fig:scalarfielddom}
\end{figure}

\begin{figure}[h]% fig.5
%\begin{center}
\includegraphics[height=5cm, width=8cm]{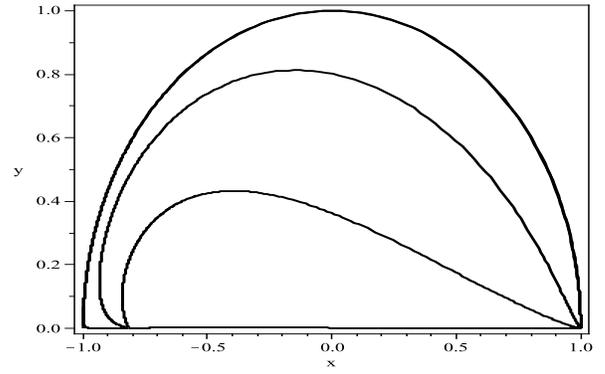}
%\end{center}
\caption{The phase plane for the potential model $V=V_0\phi^{4}$,
$\gamma=1$, and $\xi=1/4$. The attractor is a kinetic scaling
solution with $x=-\sqrt{2/3}$, $y=0$, where $\Omega_{\phi}=2/3$,
$\gamma_{\phi}=2$, $\gamma_{eff}=1/3$ and $\bar{q}= 0.5$.}
\label{fig:kineticscaling}
\end{figure}

\begin{figure}[h]% fig.6
%\begin{center}
\includegraphics[height=5cm, width=8cm]{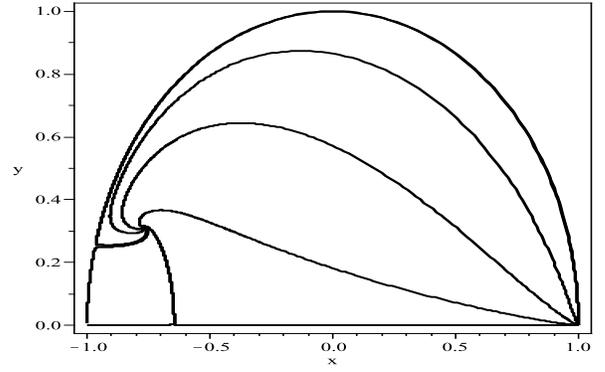}
%\end{center}
\caption{The phase plane for the potential model $V=V_0\phi^4$,
$\gamma=1$, and $\xi=2.5/16$. The scalar field dominated solution
is the saddle at $x=-\sqrt{15}/4$, $y=1/4$, and the attractor is a
scalar field scaling solution with $x=-\sqrt{3/5}$,
$y=1/\sqrt{10}$, where $\Omega_{\phi}= 7/10$,
$\gamma_{\phi}=12/7$, $\gamma_{eff}=1/2$ and $\bar{q}=1/4$.}
\label{fig:scalarfieldscaling}
\end{figure}

In this subsection, we show phase-space diagrams of the stability
solutions. Fig.~\ref{fig:phasespace} shows that different regions
in the ($\gamma$, $\xi$) parameter space lead to different
qualitative evolutions for the potential model
$V(\phi)=V_0\phi^4$.
\begin{itemize}
\item Region A
    \begin{eqnarray}
    0<\gamma\leq 4/3 \ , \qquad \xi>3(2-\gamma)^2/8 \ ,
    \end{eqnarray}
    and
        \begin{eqnarray}
    4/3<\gamma<2\ , \qquad \xi>1/6  \ .
    \end{eqnarray}
The kinetic dominated solution ($A_{-}$) is stable in this case.
The scalar field dominated solution and the kinetic scaling
solution are saddle for $\xi>1/6$ and $\xi>3(2-\gamma)^2/8$,
respectively. The scalar field scaling solution is unstable. The
effective equation of state is, accordingly,
$\gamma_{eff}=2-4(2\xi/3)^{1/2}$ and the universe accelerates for
$\xi>1/6$. For the inflationary attractor solutions, we obtain:
i)~The power-law inflationary attractor solution (Region A1)
corresponds to region of ($\gamma$, $\xi$) parameter space:
$1<\gamma<4/3$, $3(2-\gamma)^2/8<\xi<3/8$ and $4/3\leq \gamma \leq
 2$, $1/6<\xi<3/8$, ii)~The de Sitter inflationary attractor
solution (Line $\xi=3/8$) corresponds to region of ($\gamma$,
$\xi$) parameter space: $1<\gamma<2$, $\xi=3/8$, and iii)~The
superinflationary attractor solution (Region A2) correspond to
region of ($\gamma$, $\xi$) parameter space: $0 \leq\gamma\leq1$,
$\xi>3(2-\gamma)^2/8$  and $1<\gamma\leq 2$, $\xi>3/8$. In
Fig.~\ref{fig:kineticdom}, we show the phase plane plot for
$\gamma=1$ and $\xi=1$. The superinflationary attractor is the
kinetic dominated solution.
%%%%%%%%%%%%%%%%%%%%%%%%%%
\item Region B
    \begin{eqnarray}
    0<\gamma\leq 4/3 \ , \qquad 0<\xi<{\gamma\over 8} \ ,
    \label{exp4:B1}
    \end{eqnarray}
and
    \begin{eqnarray}
    4/3< \gamma\leq 2 \ , \qquad 0<\xi<1/6 \ .
        \label{exp4:B2}
    \end{eqnarray}
The scalar field dominated solution is stable in this case and it
exists for $\xi<1/6$. The kinetic dominated solution and the
scalar field scaling solution are saddle for $0<\xi<1/6$ and
$0<\xi<{\gamma/ 8}$, respectively. The kinetic scaling solution is
unstable. The universe accelerates for $0<\xi<1/6$. We only find
the power-law inflationary attractor solution in this case, and it
corresponds to region of ($\gamma$, $\xi$) parameter space:
$0<\gamma \leq 4/3$, $0<\xi<\gamma/8$ and $4/3<\gamma \leq 2$,
$0<\xi<1/6$. In Fig.~\ref{fig:scalarfielddom}, we show the phase
plane plot for $\gamma=1$ and $\xi=1/16$. The power-law
inflationary attractor is the  scalar field dominated solution.
%%%%%%%%%%%%%%%%%%%%%%%%%%
\item Region C
        \begin{eqnarray}
    0\leq\gamma<{4\over 3}  \ , \quad {3\gamma(2-\gamma)\over 16}<\xi< {3(2-\gamma)^2 \over 8} \ .
    \label{exp4:C}
    \end{eqnarray}
The kinetic scaling solution is stable in this case. The kinetic
dominated solution, the scalar field dominated solution, and the
scalar field scaling solution are saddle for $\xi< {3(2-\gamma)^2
/ 8}$, $\xi>1/6$, and ${3\gamma(2-\gamma)/16}<\xi$, respectively.
Also, for the inflationary  solutions, there are three cases to be
considered, i.e., i)~The power-law inflationary attractor solution
(Region C1) corresponds to region of ($\gamma$, $\xi$) parameter
space: $0<\gamma\leq1$,
$3\gamma(2-\gamma)/16<\xi<3\gamma(2-\gamma)/8$ and $1<
\gamma<4/3$, $3\gamma(2-\gamma)/16<\xi<3(2-\gamma)^2/8$, ii)~The
de Sitter inflationary attractor solution (Line
$\xi=3\gamma(2-\gamma)/8$) corresponds to region of ($\gamma$,
$\xi$) parameter space: $0<\gamma<1$, $\xi=3\gamma(2-\gamma)/8$,
and iii)~The superinflationary attractor solution (Region C2)
corresponds to region of ($\gamma$, $\xi$) parameter space: $0
\leq\gamma<1$, $3\gamma(2-\gamma)/8<\xi<3(2-\gamma)^2/8$.
Fig.~\ref{fig:kineticscaling} shows the phase plane plot for
$\gamma=1$ and $\xi=1/4$. The power-law inflationary attractor is
the kinetic scaling solution.
%%%%%%%%%%%%%%%%%%%%%%%%%%
\item Region D \\
The scalar field scaling solution exists for $0<\gamma<2$, $0<\xi<
3\gamma(2-\gamma)/16$, and it is stable for
\begin{eqnarray}
    0<\gamma<{4\over 3} \ , \quad {\gamma\over 8}<\xi< {3\gamma(2-\gamma)\over 16} \ .
    \label{exp4:D}
\end{eqnarray}
The kinetic dominated solution is unstable. The scalar field
dominated solution and the kinetic scaling solution are saddle for
$\xi>{\gamma/ 8}$ and $\xi< {3\gamma(2-\gamma)/6}$, respectively.
The universe accelerates for $0\leq\gamma<4/3$. For the
inflationary attractor solutions, we only find the power-law
inflationary attractor solution in this case, corresponding to the
region of ($\gamma$, $\xi$) parameter space: $0\leq\gamma < 4/3$,
$\gamma/8<\xi< 3\gamma(2-\gamma)/16$. In
Fig.~\ref{fig:scalarfieldscaling}, we show the phase plane plot
for $\gamma=1$ and $\xi=2.5/16$. The power-law inflationary
attractor is the scalar field scaling solution.
\end{itemize}
%
%===============================================================%
%************************ SECTION  *****************************%
%===============================================================%
\subsection{Cosmological implications}
As we have shown in the previous section, there are some
possibilities that lead to an accelerated expansion: the kinetic
dominated solutions and the scalar field dominated solutions
together with scaling solutions. The later, however, is not viable
to explain an accelerated universe at the present epoch since
$\Omega_\phi \simeq 0.65 \pm 0.05$ with $\omega_\phi \leq
-0.4$~\cite{Wang:2000}. In this section we consider the
cosmological implications of the scaling solutions. For the case
power-law potential, we find two scaling solutions: the kinetic
scaling solution and the scalar field scaling solution.

In the background radiation, we find the following relations for
the kinetic scaling solution:
\begin{eqnarray}
\Omega_{\phi} = 6\xi \ , \quad \gamma_{\phi}=2 \ , \quad
\gamma_{eff}= {4\over 3}(1- 3\xi)  \ . \label{cosmol:C}
\end{eqnarray}
The universe accelerates for $\xi>1/6$. However, this scaling
solution is not viable to explain the accelerated universe in the
background radiation, because the universe is unstable in this
case.

The scalar field scaling solution has the density parameter,
\begin{eqnarray}
\Omega_{\phi}={4\over n^2\xi}\left(1- {n\xi\over 2} \right) \ ,
\label{Omegascaling}
\end{eqnarray}
for $\gamma=4/3$, where ranges $\xi$  between $0<\xi<2/3n$. We
also obtain
\begin{eqnarray}
\gamma_{\phi}={8\over 3(2-n\xi)} \ , \quad \gamma_{eff}=
{4(n-2)\over 3n} \ .
\end{eqnarray}

Moreover, using the standard nucleosynthesis and the observed
abundances of primordial nuclides, the strong constraint is that
the fractional energy density of scalar field at temperatures near
1 MeV is $\Omega_\phi<\Omega_\phi^{max}$. This implies
\begin{equation}
{4 \over n(n\Omega_\phi^{max}+2)}< \xi  \ . \label{nuc1}
\end{equation}
For $\Omega_{\phi}^{max}\simeq 0.2$, the nucleosynthesis bound
requires
\begin{equation}
{20 \over n(n+10)}< \xi \ ,
 \label{nuc2}
\end{equation}
for the Lorentz violation to be relevant.
\begin{figure}[h]% fig.7
%\begin{center}
\includegraphics[height=5cm, width=8cm]{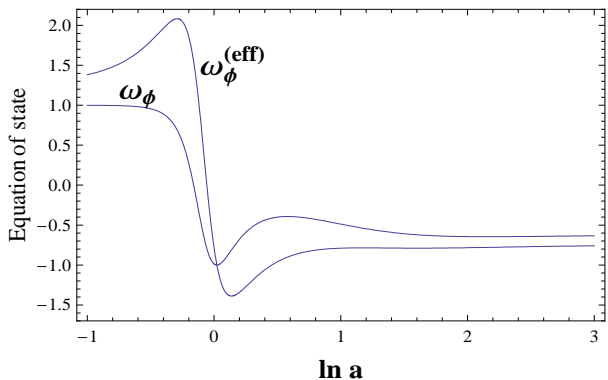}
%\end{center}
\caption{The evolutions of the effective equation of state as a
function of $\ln a$ for the case $\xi=1/32$, $\gamma=1$ and $n=4$.
We choose the initial conditions $x_i=0.01$, $y_i=0.12$.}
\label{fig:varyingeos}
\end{figure}

From Eq.~(\ref{eq:eos-mat}), we define
$\gamma_{\phi}^{(eff)}=\gamma_{\phi}+H^2
\bar{\beta}^{\prime}/\rho_{\phi}$ which is related with the
effective equation of state parameter $\omega_{\phi}^{(eff)}$ by
the relationship $\omega_{\phi}^{(eff)}=\omega_{\phi}+H^2
\bar{\beta}^{\prime}/\rho_{\phi}$. In Fig.~\ref{fig:varyingeos} we
show the evolutions of equation of state as a function of
$\alpha=\ln a$ for the case $\xi=1/32$, $\gamma=1$ and $n=4$. We
choose the initial conditions $x_i=0.01$, $y_i=0.12$. The
effective equation of state for the scalar field can cross $-1$.

%===============================================================%
%************************ CONCLUSION ***************************%
%===============================================================%
\section{Conclusions}
In this paper, we have studied the attractor solutions of the
scalar field in the frame of Lorentz violating
scalar-vector-tensor theory of gravity. In this model, because of
the dynamics of the effective coupling vector, the dynamics of the
scalar field is modified from the conventional cosmology. The
characteristic of the slope, $p$, and the condition of the
accelerating universe are given by Eqs.~(\ref{plwithmatter}) and
(\ref{acc:cond2}), respectively.

For the model (\ref{model}) without a barotropic fluid, the
critical points of a dynamical system completely depend on the
coupling parameter with typical potential. There exist two kinetic
dominated solutions and one potential-kinetic dominated solution.
Different regions in the ($\xi, p$) parameter space lead to
different qualitative evolutions. There are three typical
inflationary solutions which are obtained from
Eq.~(\ref{defpowerlaw}): the power-law inflation, the de Sitter
inflation and the superinflation. Three types of inflationary
attractor exist both in the kinetic dominated solution and scalar
field dominated solution. Note that the superinflationary is
stable in the case purely Lorentz violation without potential,
where $\xi>3/2$.

We have extended the dynamical system of the evolution to a
realistic universe model with a barotropic fluid. The stability of
the critical point was discussed in Sec. IV. In this case, the
fluid dominated solution is absent, instead, we find two scaling
solutions: the kinetic scaling solution and the scalar field
scaling solution. We have completely have shown the stable
attractor in regions of ($\gamma$, $\xi$) parameter space. Based
on this parameter space, we have shown the phase plane plot for
the four different qualitative evolutions, i.e. the kinetic
dominated solution, the scalar field (potential-kinetic) dominated
solution, the kinetic scaling solution and the scalar field
scaling solution. Another result is that the scalar-vector-tensor
theory of gravity is still relevant at the nucleosynthesis. From
this data, we derive the constraints for the value of the coupling
parameter in Eqs.~(\ref{nuc1}).

More interestingly, the model (\ref{model}) allows for the
inverse-power potential, $n<0$. As an example, in the case $n=-2$,
we only have one attractor solution in the potential dominated
solution. The effective equation of state parameter
$\gamma_{\phi}$ is of the cosmological constant. Two attractor
solutions, the scalar field dominated solution together with the
scaling solution exist for $n<-2$.

When $\lambda_1$ is dynamically changing quantity, the effective
coupling vector could track the evolution of the scalar field.
There exists one stable critical point that gives an acceleration
of the universe at late time with the equation of state parameter
$\omega <-1/3$. Thus, we can have a dark energy scenario in the
frame of scalar-vector-tensor theory of gravity \cite{zen:2007}.
Finally, we would like to emphasize that there exists an attractor
solution in the Lorentz violating scalar-vector-tensor theory of
gravity.

\begin{acknowledgements}
Arianto wishes to acknowledge all members of the Theoretical
Physics Laboratory, the THEPI Divison of the Faculty of
Mathematics and Natural Sciences, ITB, for the warmest
hospitality. This research is financially supported by Riset
Internasional ITB, No. 054/K01.7/PL/2008.
\end{acknowledgements}

\appendix*
\begin{table*}
\caption{\label{tab:table1}The properties of critical points with
barotropic fluid in the Lorentz violating scalar-vector-tensor
theory of gravity for constants $\lambda_1$ and  $\lambda_2$. }
\begin{ruledtabular}
\begin{tabular}{|c|c|c|c|c|}
 Point&$x$&$y$ &Existence&Stability  \\ \hline
 $(A_{+})$&$+1$&$0$ &All $\lambda_1$, $\lambda_2$ and $\gamma$&Stable: $ \lambda_1 >{6-\sqrt{6}\lambda_2}/{\sqrt{6}}$  and $\lambda_1 > {3(2-\gamma)}/{\sqrt{6}}$ \\ \hline

 $(A_{-})$&$-1$&$0$ &All $\lambda_1$, $\lambda_2$  and $\gamma$&Stable: $\lambda_1 <-(6+\sqrt{6}\lambda_2)/{\sqrt{6}}$ and $\lambda_1 <  -3(2-\gamma)/{\sqrt{6}}$\\  \hline

 $(B)$   &${1\over \sqrt{6}}\left(\lambda_1+ \lambda_2\right)$&$\left[1 - {(\lambda_1 + \lambda_2)^2 \over 6} \right]^{1/2}$&$(\lambda_1+\lambda_2)^2<6$&Stable: See Eq. (\ref{app:stabilityB}) \\  \hline

 $(C)$   &$(2/3)^{1/2} {\lambda_1 \over (2-\gamma)}$&$0$ &All $\lambda_2$, $\gamma \neq 2$ and $\lambda_1\neq 0$&Stable: See Eq. (\ref{app:stabilityC}) \\  \hline

 $(D)$   &$(3/2)^{1/2}{\gamma\over \lambda_2}$&$\left[{3\gamma(2-\gamma)\over 2\lambda_2^2}-{\lambda_1\over \lambda_2} \right]^{1/2}$ &$\lambda_1\lambda_2<{3\gamma(2-\gamma)\over 2}$&Stable: See Eqs. (\ref{app:stabilityD1})-(\ref{app:stabilityD2}) \\
\end{tabular}
\end{ruledtabular}
\end{table*}
\begin{table*}
\caption{\label{tab:table2}The scalar field density parameter
$\Omega_\phi$, the equation of state $\gamma_\phi$, the effective
equation of state $\gamma_{eff}$, the slope $p$ and  the
acceleration parameter, $\bar{q}$, for the cases of constants
$\lambda_1$ and  $\lambda_2$.}
\begin{ruledtabular}
\begin{tabular}{|c|c|c|c|c|c|}
Point  &  $\Omega_{\phi}$ & $\gamma_{\phi}$ & $\gamma_{eff}$ & $1/p$ & $\bar{q}$\\
\hline
$(A_{+})$  &  1 & 2 & $2-2\sqrt{2/3}\lambda_1$ & $3-\sqrt{6}\lambda_1$ & $-2+\sqrt{6}\lambda_1$ \\
\hline

$(A_{-})$  &  1 & 2 & $2+2\sqrt{2/3}\lambda_1$ & $3+\sqrt{6}\lambda_1$ & $-2-\sqrt{6}\lambda_1$ \\
\hline

$(B)$  &  1 & ${1\over 3}\left(\lambda_1+ \lambda_2 \right)^2$ & ${1\over 3}\left(\lambda_2^2-\lambda_1^2\right)$ & ${1\over 2}(\lambda_2^2-\lambda_1^2)$ & $1-{1\over 2}(\lambda_2^2-\lambda_1^2)$ \\
\hline

$(C)$  &  ${2\lambda_1^2 \over 3(2-\gamma)^2}$ & 2 & $\gamma-{2\lambda_1^2\over 3(2-\gamma)}$ & ${3\over 2}\gamma- {\lambda_1^2\over (2-\gamma)}$ & $1-{3\over 2}\gamma+ {\lambda_1^2\over (2-\gamma)}$ \\
\hline

$(D)$  &  ${3\gamma \over \lambda_2^2}\left(1-{\lambda_1\lambda_2\over 3\gamma} \right)$ & $\gamma\left(1-{\lambda_1\lambda_2 \over 3\gamma } \right)^{-1}$ & $\gamma\left(1-{\lambda_1\over \lambda_2} \right)$ & ${3\over 2}\gamma\left(1-{\lambda_1\over \lambda_2} \right)$ & $1-{3\over 2}\gamma\left(1-{\lambda_1\over \lambda_2} \right)$ \\
\end{tabular}
\end{ruledtabular}
\end{table*}
\section{Stability of the critical points for constants $\lambda_1$ and  $\lambda_2$}
Let us consider the cases $\Gamma_1=1/2$ and $\Gamma_2=1$. From
Eqs.~(\ref{auto-L1})--(\ref{auto-L2}), we obtain $\lambda_1$ and
$\lambda_2$ nearly constants. The critical points $(x_c, y_c)$ are
obtained by imposing the conditions $x^{\prime}=0$ and
$y^{\prime}=0$. Substituting linear perturbation $x \rightarrow
x_c+u$ and $y \rightarrow y_c+v$ about the critical points into
Eqs.~(\ref{autobarotropic-x}) and (\ref{autobarotropic-y}), to
first-order in the perturbation, one obtains the equations of
motion
\begin{eqnarray}
\label{matrix} \left(
  \begin{array}{c}
    u^{\prime} \\
    v^{\prime} \\
  \end{array}
\right)= M\left(
  \begin{array}{c}
    u \\
    v \\
  \end{array}
\right) \ ,
\end{eqnarray}
where the elements of the matrix $M$ are
\begin{eqnarray}
M_{11}&=&  -3+3(2-\gamma)x^2+{3\over 2}\left[\gamma\left(1-x^2-y^2\right)+2x^2\right] \nonumber\\
&&-\sqrt{6}\lambda_1 x \ ,
\label{M11}\\
M_{12}&=& \left(\sqrt{6} \lambda_2 -3\gamma x\right)y \ ,
\label{M12}\\
M_{21}&=& \left[3(2-\gamma)x - \sqrt{{3\over
2}}(\lambda_1+\lambda_2) \right]y \ ,
\label{M21}\\
M_{22}&=& -3\gamma y^2+{3\over 2}\left[\gamma\left(1-x^2-y^2\right)+2x^2\right] \nonumber\\
&&- \sqrt{{3\over 2}}(\lambda_1+\lambda_2)  x \ . \label{M22}
\end{eqnarray}

The critical points together with the stability analysis for
constants $\lambda_1$ and $\lambda_2$ are shown in
Table~\ref{tab:table1}. For completeness, the scalar field density
parameter $\Omega_\phi$, the equation of state $\gamma_\phi$, the
effective equation of state $\gamma_{eff}$, the slope $p$, and the
acceleration parameter $\bar{q}$ are shown in
Table~\ref{tab:table2}. The eigenvalues of the stability matrix
$M$, Eq.~(\ref{matrix}), are as follows:
\begin{itemize}
    \item Point $(A_{\pm})$\\ $(x_c,y_c)=(\pm 1,0)$ is a kinetic dominated solution,
    \begin{eqnarray}
    m_1&=&3 \mp \sqrt{{3\over 2}} \left(\lambda_1+ \lambda_2\right) \ ,
    \label{ev:Apm-1} \\
    m_2&=&3(2-\gamma) \mp \sqrt{6}\lambda_1 \ .
    \label{ev:Apm-2}
    \end{eqnarray}
The fixed point $(A_{+})$ is stable for
    \begin{eqnarray}
     \lambda_1 >\frac{6-\sqrt{6}\lambda_2}{\sqrt{6}} \ , \quad\textrm{and} \quad \lambda_1 >  \frac{3(2-\gamma)}{\sqrt{6}} \ ,
    \end{eqnarray}
and the fixed point $(A_{-})$ is stable for
    \begin{eqnarray}
   \lambda_1 <-\frac{6+\sqrt{6}\lambda_2}{\sqrt{6}} \ , \quad\textrm{and} \quad \lambda_1 <  -\frac{3(2-\gamma)}{\sqrt{6}} \ .
    \end{eqnarray}
    \item Point $(B)$\\
    ($x_c,y_c$)=($(\lambda_1+\lambda_2)/\sqrt{6},[1-(\lambda_1+\lambda_2)^2/6]^{1/2}$) is a scalar field dominated solution,
     \begin{eqnarray}
    m_1&=& -3+{1 \over 2}\left(\lambda_1+ \lambda_2\right)^2 \ ,
    \label{ev:Bm1}\\
    m_2&=& -3\gamma +\lambda_2\left(\lambda_1+ \lambda_2\right) \ .
    \label{ev:Bm2}
    \end{eqnarray}
The fixed point is a stable for
     \begin{eqnarray}
     \label{app:stabilityB}
     &&-\sqrt{6}-\lambda_2<\lambda_1< \sqrt{6}-\lambda_2 \ , \quad\textrm{and}\nonumber\\
     && \lambda_1> \frac{3\gamma-\lambda_2^2}{\lambda_2}  \quad\textrm{for} \quad \lambda_2<0 \ , \quad\textrm{or}\nonumber\\
     && \lambda_1< \frac{3\gamma-\lambda_2^2}{\lambda_2}  \quad\textrm{for} \quad \lambda_2>0 \ ,
    \end{eqnarray}
    \item Point $(C)$ \\
    ($x_c,y_c$)=($(2/3)^{1/2}\lambda_1/(2-\gamma),0$) is a kinetic scaling solution,
    \begin{eqnarray}
    m_1&=& -{3\over 2}(2-\gamma) +{\lambda_1^2 \over 2-\gamma} \ ,
    \label{ev:C}\\
    m_2&=& {3\gamma \over 2}-{\lambda_1\lambda_2\over 2-\gamma} \ .
    \end{eqnarray}
The fixed point is stable for
    \begin{eqnarray}
    &&\lambda_1^2 > {3\over 2}(2-\gamma)^2 \ , \qquad\textrm{and}\nonumber\\
    && \lambda_1 > \frac{3\gamma(2-\gamma)}{2\lambda_2} \quad\textrm{for} \quad \lambda_2<0 \ ,\quad\textrm{or}\nonumber\\
    && \lambda_1 < \frac{3\gamma(2-\gamma)}{2\lambda_2} \quad\textrm{for} \quad \lambda_2>0 \ .
    \label{app:stabilityC}
    \end{eqnarray}
    \item Point $(D)$\\
    ($x_c,y_c$)=($(3/2)^{1/2}{\gamma/ \lambda_2}$,$\left[{3\gamma(2-\gamma)/ 2\lambda_2^2}-{\lambda_1/ \lambda_2} \right]^{1/2}$) is a scalar field scaling solution,
 \end{itemize}
    \begin{eqnarray}
   && m_{1,2}=-{3\over 4}\left[2-\gamma\left(1 + {\lambda_1\over \lambda_2}\right) \right]  \times \nonumber\\
    &&\left[1\pm \sqrt{1-\frac{8(3\gamma^2-6\gamma+2\lambda_1\lambda_2)(3\gamma-\lambda_1\lambda_2-\lambda_2^2)}{3\lambda_2^2\left[2-\gamma\left(1 + {\lambda_1\over \lambda_2}\right)\right]^2}} \right] \ . \nonumber\\
    \label{ev:D}
    \end{eqnarray}
The fixed point is a stable for
    \begin{eqnarray}
    &&\lambda_1<\frac{2-\gamma}{\gamma}\lambda_2 \ ,\qquad\textrm{and} \nonumber\\
    &&\lambda_1<\frac{3\gamma(2-\gamma)}{2\lambda_2} \ ,\qquad \lambda_1 >\frac{3\gamma-\lambda_2^2}{\lambda_2} \ ,
    \label{app:stabilityD1}
    \end{eqnarray}
where $\lambda_2<0$,
     \begin{eqnarray}
    &&\lambda_1<\frac{2-\gamma}{\gamma}\lambda_2 \ ,\qquad\textrm{and} \nonumber\\
    &&\lambda_1>\frac{3\gamma(2-\gamma)}{2\lambda_2} \ ,\qquad \lambda_1 <\frac{3\gamma-\lambda_2^2}{\lambda_2} \ ,
    \label{app:stabilityD2}
    \end{eqnarray}
where $\lambda_2>0$.

\end{document}